\documentclass[aps,prd,floatfix,superscriptaddress,preprintnumbers,nofootinbib,twocolumn,longbibliography]{revtex4-1}
\pdfoutput=1

\usepackage{amsmath, amssymb, amsfonts, graphicx, mathrsfs, verbatim, float, slashed, bbm, color, xcolor, xspace, enumerate, url, bbding, multirow, outlines,upgreek}
\usepackage[english]{babel}
\usepackage[colorlinks=true,linkcolor=xenon1teal,citecolor=xenon1teal]{hyperref}
\definecolor{xenon1teal}{rgb}{0.3,0.7, 0.8}

\begin{document}

\title{Electric But Not Eclectic: Thermal Relic Dark Matter for the XENON1T Excess}

\author{Joseph Bramante}
\affiliation{The McDonald Institute and Department of Physics, Engineering Physics, and Astronomy, Queen's University, Kingston, Ontario, K7L 2S8, Canada}
\affiliation{Perimeter Institute for Theoretical Physics, Waterloo, Ontario, N2L 2Y5, Canada}

\author{Ningqiang Song}
\affiliation{The McDonald Institute and Department of Physics, Engineering Physics, and Astronomy, Queen's University, Kingston, Ontario, K7L 2S8, Canada}
\affiliation{Perimeter Institute for Theoretical Physics, Waterloo, Ontario, N2L 2Y5, Canada}

\begin{abstract}
The identity of dark matter is being sought with increasingly sensitive and voluminous underground detectors. Recently the XENON1T collaboration reported excess electronic recoil events, with most of these having recoil energies around $1-30$ keV. We show that a straightforward model of inelastic dark matter produced via early universe thermal freeze-out annihilation can account for the XENON1T excess. Remarkably, this dark matter model consists of a few simple elements: sub-GeV mass Dirac fermion dark matter coupled to a lighter dark photon kinetically mixed with the Standard Model photon. A scalar field charged under the dark U(1) gauge symmetry can provide a mass for the dark photon and splits the Dirac fermion component state masses by a few keV, which survive in equal abundance and interact inelastically with electrons and nuclei.
\end{abstract}

\maketitle

\section{Introduction}
While ample evidence has been collected demonstrating the gravitational influence dark matter (DM) exerts on galaxies and structure formation in the early universe, DM's origin, couplings, and mass remain a compelling mystery. If DM is a particle with a mass less than a gram, then the predicted flux of DM at Earth's position implies that DM's interactions could be detected with multi-tonne-scale detectors, although this will depend on its coupling to known particles. At present some of the most incisive searches for DM are being conducted in low-background laboratories deep underground.

Recently the XENON1T collaboration reported an excess of electron recoil events in a 0.65 tonne-year exposure of cooled xenon, with many events having recoil energies around a few keV. Since this xenon search is the most voluminous and sensitive search ever conducted at keV recoil energies, it is possible these excess events are attributable to a hitherto undetected background process: the beta decay of tritium has been proposed as one such background \cite{Aprile:2020tmw}. However by the same token it is possible that XENON1T has discovered the interactions of a DM particle. Since the XENON1T result was announced, a number of new physics proposals have been put forth to explain the excess \cite{Aprile:2020tmw,Smirnov:2020zwf,Takahashi:2020bpq,Kannike:2020agf,Fornal:2020npv,Boehm:2020ltd,Alonso-Alvarez:2020cdv,Buch:2020mrg,DiLuzio:2020jjp,Paz:2020pbc,Bell:2020bes,AristizabalSierra:2020edu,Chen:2020gcl,Choi:2020udy,Harigaya:2020ckz,Lee:2020wmh,Primulando:2020rdk,Nakayama:2020ikz,Khan:2020vaf,Dey:2020sai,Su:2020zny,Du:2020ybt,Bally:2020yid,Cao:2020bwd,Dey:2020sai}. However, thus far it has appeared difficult to explain the excess without invoking special DM or dark sector (DS) properties.

Here we will demonstrate that a straightforward model of Dirac fermion dark matter, coupled to the Standard Model (SM) through a dark photon, can account for the observed DM relic abundance and may have been detected as an excess of electron recoil events at XENON1T. Two key features of this model are an inelastic mass splitting of a few keV between the Dirac fermion component states and that the DM mass is greater than the dark photon mass, so that annihilation of DM in the early universe proceeds predominantly through annihilation to dark photons. As we will see, the XENON1T excess can be accounted for by inelastic down-scatters depositing a few keV of energy into electrons at XENON1T.

\section{Inelastic Dark Photon Mediated Dark Matter}

The kinematics and characteristics of inelastic DM models have been studied extensively \cite{Hall:1997ah,TuckerSmith:2001hy,TuckerSmith:2004jv,Finkbeiner:2007kk,Pospelov:2007mp,Chang:2008gd,ArkaniHamed:2008qn,Alves:2009nf,Lisanti:2009am,Cheung:2009qd,Cui:2009xq,Batell:2009vb,Fox:2010bu,Graham:2010ca,Essig:2010ye,Pospelov:2013nea,Dienes:2014via,Barello:2014uda,Izaguirre:2015yja,Bramante:2016rdh,Darme:2017glc,Darme:2018jmx}. Inelastic DM mediated by a dark photon has been examined in $e.g.$ \cite{Batell:2009vb,Izaguirre:2015yja,Bramante:2016rdh}. Hereafter our conventions and treatment will follow Reference \cite{Bramante:2016rdh} most closely, although there are some key differences, since \cite{Bramante:2016rdh} primarily focused on DM masses in excess of 100 GeV, while here we find some details are different for sub-GeV mass DM that explains the XENON1T excess.

We consider a massive dark photon $V$, Dirac fermion $\psi$, and complex scalar $\phi$, all charged under a $U(1)_D$ gauge symmetry. The Lagrangian is
\begin{align}
\mathcal{L}  =  ~&\mathcal{L}_{ SM} + |D_{\mu} \phi|^2 - V(\phi) - \frac{1}{4} V_{\mu \nu}V^{\mu \nu} + \epsilon V_\mu \partial_\nu F^{\mu \nu} \nonumber \\  &+ \bar{\psi} (iD_\mu \gamma^\mu - m_{\psi})\psi +(y_{D} \phi\, \bar{\psi}^T\, C^{-1}\, \psi + {\rm h.c.}),
\label{eq:dplag}
\end{align}
where $D_\mu \equiv  \partial_\mu + i g_D V_\mu$ is the gauge covariant derivative with gauge coupling $\alpha_D \equiv g_D^2/4 \pi$, $V_{\mu \nu}$ and $F_{\mu \nu}$ are the dark and SM field strength tensors, $C$ is the charge conjugation matrix for $\psi$, and $y_D$ is the Yukawa coupling between $\phi$ and $\psi$. $V$ can obtain a mass term of the form $M_V^2 V_\mu V^\mu$ either through the Stueckelberg mechanism or through coupling to $\phi$. We assume that $\phi$ obtains a vacuum expectation value (vev) $v_\phi$ through the machinations of its potential $V(\phi)$. Then the Dirac fermion component mass states, which we label $\chi_{2,1}$ will be split by a mass difference
\begin{equation}
    \delta \equiv M_{\chi_2} - M_{\chi_1} \simeq y_D v_\phi = {\rm keV} ~\left(\frac{y_d}{10^{-7}} \right)\left(\frac{v_\phi}{10~{\rm GeV}} \right),
\end{equation}
where here we have normalized $\delta \sim{\rm keV}$ which will match the XENON1T excess, and the scalar vev to a value which would permit $v_\phi$ to generate a sub-GeV mass for $V$, in a DM thermal freeze-out model.

\section{Cosmological production}
In the early Universe the dark sector will be in thermal equilibrium with the SM plasma. Freeze-out of $\chi_1$ and $\chi_2$ takes place when the temperature of the Universe drops below $M_\chi$. We are interested in the ``secluded'' DM scenario where $M_V < M_\chi$, so that the annihilations of $\chi_1$ and $\chi_2$ are dominated by the process $\bar{\chi}\chi\rightarrow VV$. This annihilation cross section is
\begin{equation}
    \sigma v=\dfrac{\pi \alpha_D^2}{M_\chi^2}\sqrt{1-\dfrac{M_V^2}{M_\chi^2}}\,.
    \label{eq:anni}
\end{equation}
To find the DM relic abundance from freeze-out annihilation, we use the standard formula \cite{Bramante:2017obj,Kolb:1990vq}
\begin{equation}
    \Omega_x h^2 =  \frac{10^9 x_f}{\sqrt{g_*} M_{Pl} \left \langle \sigma v \right\rangle ~{\rm GeV}} \approx 0.12,
    \label{eq:relicab}
\end{equation}
where $\Omega_x h^2$ is the comoving relic abundance of DM, $g_* \sim 10$ is the number of relativistic degrees of freedom at the time sub-GeV mass DM falls out of equilibrium, $x_f = M_\chi / T_f \sim 20$ is the mass-normalized freeze-out temperature, and $M_{Pl}$ is the Planck mass. Using the  $\bar{\chi}\chi\rightarrow VV$ annihilation cross-section in this relic abundance formula, we find the dark gauge coupling that satisfies DM relic abundance requirements,
\begin{align}
    \alpha_D \simeq 4 \times 10^{-5} \left( \frac{M_\chi}{\rm GeV} \right) \left(1-\frac{M_V^2}{M_\chi^2} \right)^{-1/4}.
    \label{eq:adrelic}
\end{align}
The above treatment of $\chi$'s relic abundance has neglected the possible effect of $\chi_{1,2}$ mass splitting $\delta$ on thermal freeze-out. This is warranted, since $\delta  \ll T_f$, and so the mass splitting shouldn't affect freeze-out. 

After freeze-out, the inter-conversion process $\chi_2\chi_2\leftrightarrow\chi_1\chi_1$ will be efficient until the temperature of the dark sector drops below some temperature $T_{co}$. If $T_{co} < \delta$, the ratio of the number density of $\chi_2$ and $\chi_1$ is exponentially suppressed
$
    n_2/n_1\sim e^{-\frac{\delta}{T_{co}}}\,,
$
where the inter-conversion ceases at
\begin{equation}
    \dfrac{n_2\langle\sigma_{\chi_2\chi_2\rightarrow\chi_1\chi_1}v\rangle}{H}\sim 1\,.
    \label{eq:conversion}
\end{equation}
However, in our model we note that the temperature of the dark sector drops rapidly after decoupling from electrons in the thermal bath at $T_{de}$, since after this \mbox{$T^2 \sim T_D T_{de}$}, where $T_D$ is the DM temperature. In fact, we find that inter-conversion shuts off at \mbox{$T_{de} > T_{co}\gg$~keV}, and so $n_2/n_1 \sim 1$. We estimate $T_{co}$ as follows: after freeze-out $\chi$ will be non-relativistic and the inter-conversion cross section $\sigma_{\chi_2\chi_2\rightarrow\chi_1\chi_1} \sim \alpha_D^2M_\chi^2/M_V^4$. The $\chi_2$ number density is $n_2\sim \frac{T^3 {\rm eV}}{M_\chi}$ for a matter-radiation equality temperature $T\sim {\rm eV}$, and the DM velocity $v\sim \sqrt{T_D/M_\chi}$. The conversion rate is compared to Hubble $H\sim T^2/M_{Pl}$. From Eq.~\eqref{eq:conversion} it follows that
\begin{equation}
    T_{co}\sim \frac{M_V^4}{\alpha_D^2 M_\chi^{1/2} T_{de}^{1/2}\,{\rm eV}\,M_{Pl}}\,.
\end{equation}

For a DM mass $M_\chi = 1$~GeV, a mediator mass $M_V = 0.1$~GeV, $\alpha_D = 4\times 10^{-5}$, and using an electron kinetic decoupling $T_{de} \sim {\rm MeV}$ (found using similar Hubble rate matching arguments), we find $T_{co}\sim 100~\mathrm{keV}\gg \delta\sim \mathrm{keV}$. This estimate only represents a lower limit on $T_{co}$. In most of our parameter space, inter-conversion will cease at temperatures above 100~keV. The same estimate can be applied to other DM and mediator masses, and we find $T_{co}>\delta$ for DM models explaining the XENON1T excess. Consequently we take $n_2=n_1$ in our analysis.

After freeze-out $\chi_2$ may decay to $\chi_1$ and SM particles. Since $\delta<2m_e$, $\chi_2$
may only decay to neutrinos and photons. In the presence of V-Z mixing, the $\chi_2\rightarrow \chi_1\bar{\nu}\nu$ decay rate is given by~\cite{Batell:2009vb}
\begin{equation}
    \Gamma_{\chi_2\rightarrow \chi_1\bar{\nu}\nu}=\dfrac{4\sin^2\theta_W^4}{315\pi^3}\dfrac{G_F^2\delta^9}{M_V^4}\dfrac{\epsilon^2\alpha_D}{\alpha}\,.
\end{equation}
We require the lifetime of $\chi_2$ to be longer than the age of universe in order for $\chi_2$ to be stable, which gives
\begin{equation}
    \epsilon<0.007\sqrt{\dfrac{\alpha}{\alpha_D}}\left(\dfrac{\mathrm{MeV}}{\delta }\right)^{9/2}\left(\dfrac{M_V}{100~\mathrm{MeV}}\right)^2\,.
\end{equation}
We will be particularly interested in a mass splitting $\delta\sim 3$~keV, where a decay rate suppression factor of $10^{23}$ is expected relative to the normalization given above, and there is no meaningful constraint on $\epsilon$. We conclude that for parameters around $M_\chi\sim 1$~GeV, $M_V\sim 0.1$~GeV and $\alpha_D\sim 4\times 10^{-5}$, $\epsilon$ is not constrained by decay to neutrinos. While $\chi_2$ can also decay to $\chi_1$ via the emission of three photons $\chi_2\rightarrow \chi_1+3\gamma$, the decay rate in this case is even more suppressed: $\Gamma \propto (\delta/\mathrm{MeV})^{13}$~\cite{Batell:2009vb}. Therefore we conclude $\chi_2$ is stable for the DM, dark photon, and $\delta$ masses we are interested in.

Lastly, we address $V$ decay. For sub-GeV mass DM, there are bounds on $\chi_{1,2}$ annihilation to SM particles from distortion of the cosmic microwave background (CMB) \cite{Slatyer:2009yq}. However in our setup, $\chi_{1,2}$ annihilate overwhelmingly to $VV$, and so CMB bounds do not apply unless $V$ decays mostly to SM particles. Currently the CMB bound \cite{Aghanim:2018eyx} requires that the branching fraction of $V$ to SM particles versus DS particles satisfies $\Gamma_{V \rightarrow SM} \lesssim 10^{-2} ~(M_\chi/{\rm GeV}) ~\Gamma_{V \rightarrow DS}$. 

In this paper, we give one example that satisfies the $V$ invisible decay requirement, by adding a less massive, but otherwise identical extra dark photon and fermion ($V_E,\chi_E$) to our Lagrangian \eqref{eq:dplag}, where $\chi_E$ is charged under both groups $U(1)_D \times U(1)_E$, with couplings $\alpha_D$ and $\alpha_E$. In this model, $\chi_E$ serves as the invisible decay product of $V$, while $V_E$ ensures $\chi_E$ freezes out of equilibrium to a negligible relic abundance, through the process $\bar \chi_E \chi_E \rightarrow V_E V_E $, after which $V_E$ decays to SM particles through a photon kinetic mixing of size $\epsilon_E$.  Crucially, the primary DM field $\chi$ is still only charged under $U(1)_D$ with coupling $\alpha_D$, so Eq.~\eqref{eq:adrelic} still fixes $\chi$'s relic abundance. Then if we require $m_V > 2 m_{\chi_E} > 2m_{V_E}$, $V$ will decay mostly to $\chi_E$, and the invisible $V$ decay requirement can be easily satisfied. This can be verified by considering \mbox{$\frac{\Gamma_{V \rightarrow SM}}{\Gamma_{V \rightarrow DS}} \sim \frac{M_V \alpha \epsilon^2}{M_V \alpha_D} \sim 10^{-4} ~\left( \frac{ \epsilon^2}{10^{-6}} \right) ~\left( \frac{ 4 \times 10^{-5}}{\alpha_D} \right)$}. Finally, we note that so long as ($V_E,\chi_E$) have masses greater than 10 MeV, constraints on $\Delta N_{eff}$ at recombination and big bang nucleosynthesis can be satisfied \cite{Nollett:2013pwa,Boehm:2013jpa,Fradette:2014sza}, although CMB Stage IV searches may become sensitive to $V_E$ as light as 17 MeV \cite{Ibe:2019gpv}. As a benchmark point, we consider \mbox{$M_{\chi_E} = 25~{\rm MeV}$}, $\alpha_E = 0.01$, $M_{V_E} = 20~{\rm MeV}$, and $\epsilon_E = 10^{-4}$, which satisfies these $\Delta N_{eff}$ bounds. The benchmark also yields a small freeze-out $\chi_E$ abundance $\Omega_{\chi_E} / \Omega_x \sim 10^{-8}$, $c.f.$ Eq.~\eqref{eq:adrelic}, which avoids the CMB-era ($t_{re} \sim 1~{\rm Myr}, z \sim 600$) annihilation bound \cite{Aghanim:2018eyx}, which is weakened by a factor $(\Omega_{\chi_E} / \Omega_x)^2$. Similarly, the $\chi_E$ produced from later CMB-era $\chi \chi \rightarrow \chi_E \chi_E$ annihilations are too under-abundant to be constrained by Planck. An ``on-the-spot" \cite{Slatyer:2009yq} $\chi_E$ number density computation \mbox{$n_{\chi_E} \sim 2 n_{\chi}^2 \left\langle \sigma_{\chi \chi} v \right\rangle  (1+z)^3 t_{re}$} indicates a tiny abundance from $\chi$ annihilation, $n_{\chi_E}^{CMB} / n_\chi^{CMB} \sim  10^{-9} (\rm GeV/M_\chi) $.

\section{Downscattering From Dark Photon Dark Matter}

\begin{figure}[ht!]
    \centering
    \includegraphics[width=0.9\columnwidth]{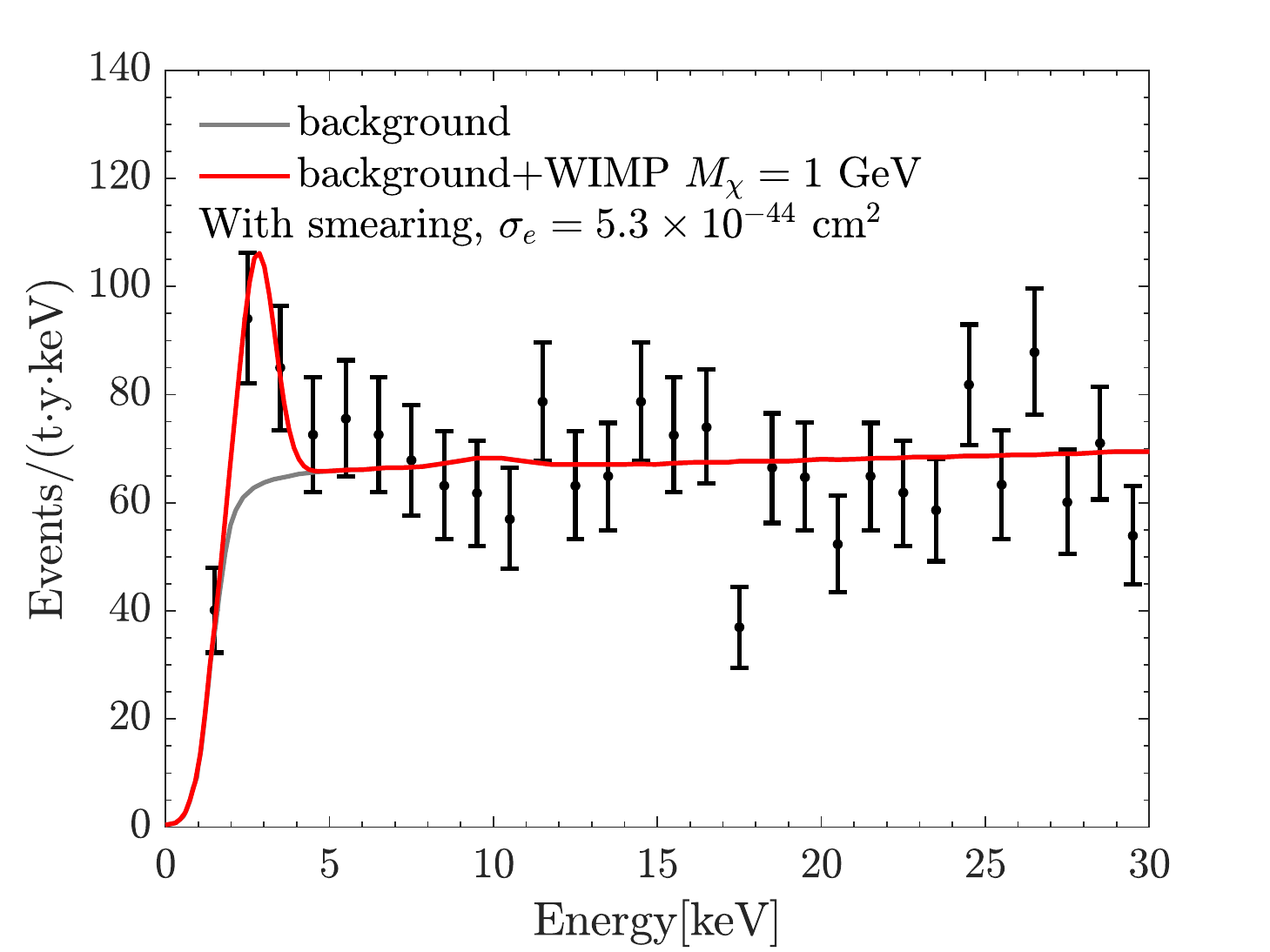}
    \includegraphics[width=0.9\columnwidth]{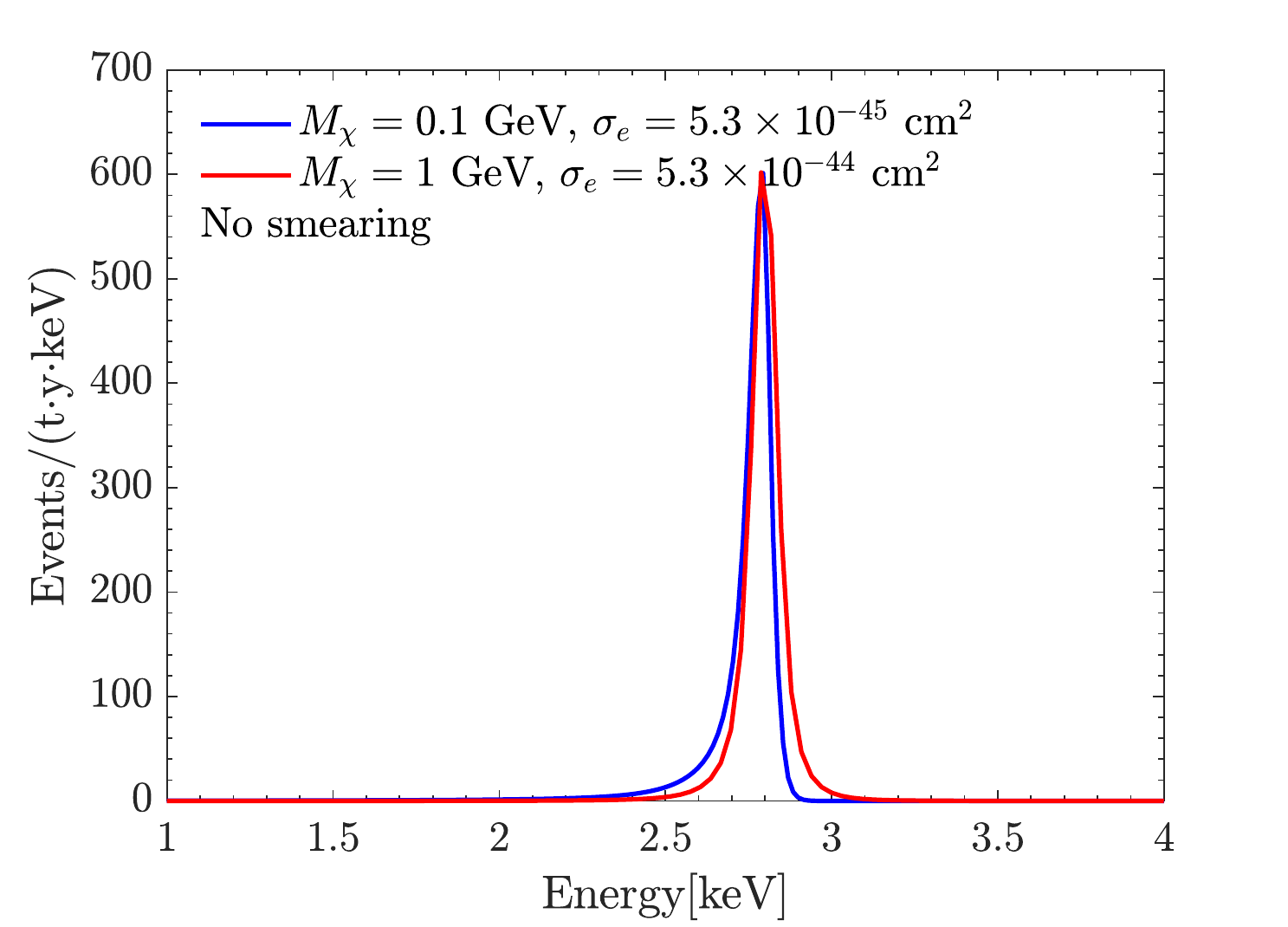}
       \includegraphics[width=0.9\columnwidth]{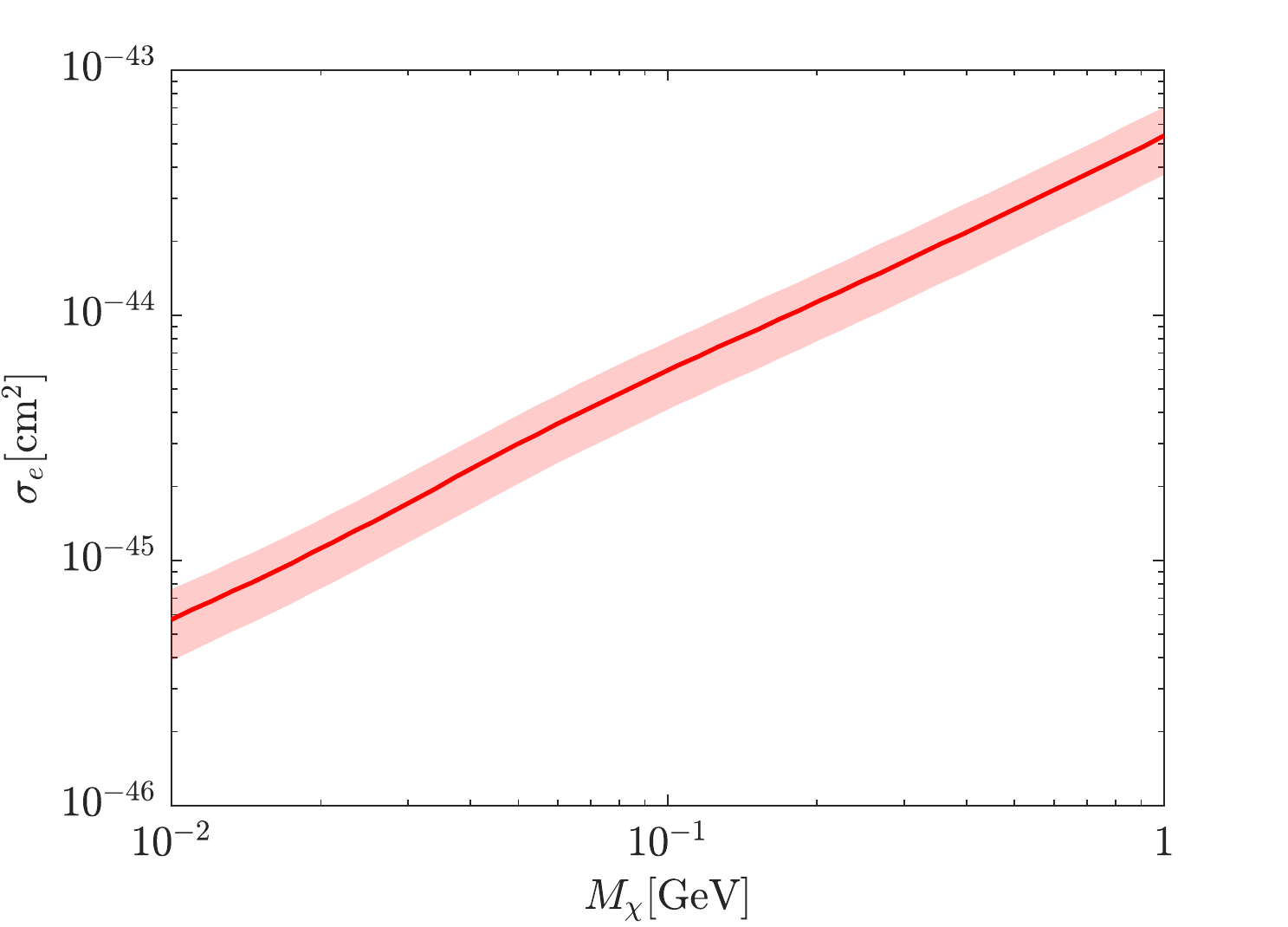}
    \caption{Exothermic DM-electron scattering fit to the XENON1T excess. Upper: Event rates including both background and exothermic scattering with $M_\chi=1$~GeV, $\delta =2.8$~keV and ${\sigma}_e=4.5\times10^{-44}$ cm$^2$. We assume $\rho_{\chi_2}=\rho_\mathrm{DM}/2$ and the detector resolution smearing has been incorporated appropriately. The Xenon1T data and background are extracted from~\cite{Aprile:2020tmw}. Middle: Exothermic scattering rates without background and detector smearing. Red and blue lines correspond to $M_\chi=1$~GeV and 0.1~GeV respectively, for $\sigma_e$ indicated. Lower: The best-fit exothermic DM-electron scattering cross-section is given over a range of DM masses, matching the XENON1T electron recoil data. The band shows the 1$\sigma$ preferred region, for $\delta =2.8$~keV, and $\rho_{\chi_2}=\rho_\mathrm{DM}/2$.}
    \label{fig:eventrate}
\end{figure}

\begin{figure*}[ht!] 
\centering
 \includegraphics[width=0.85\columnwidth]{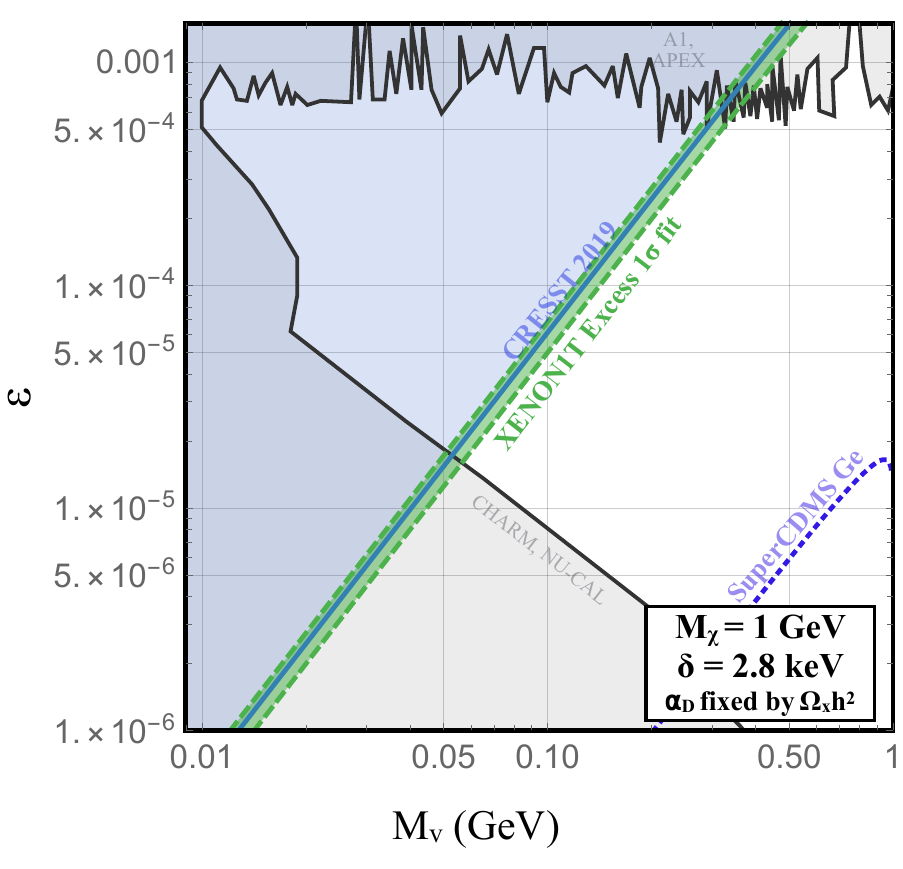}~~~~~~
 \includegraphics[width=0.85\columnwidth]{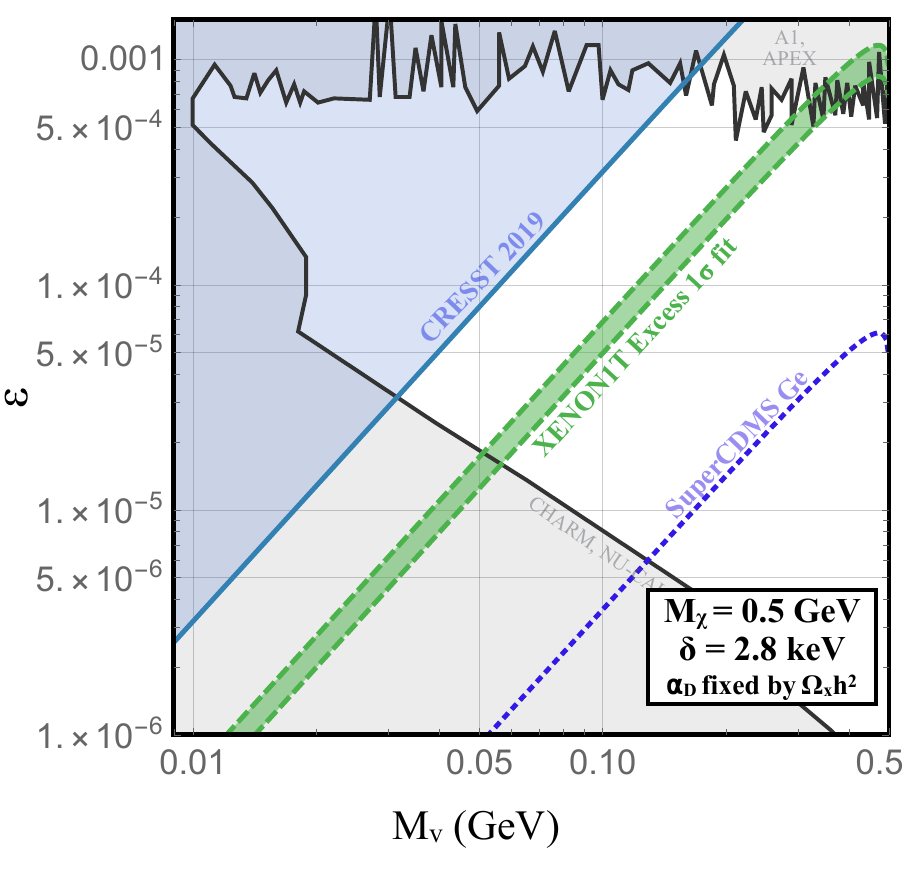}
  \includegraphics[width=0.85\columnwidth]{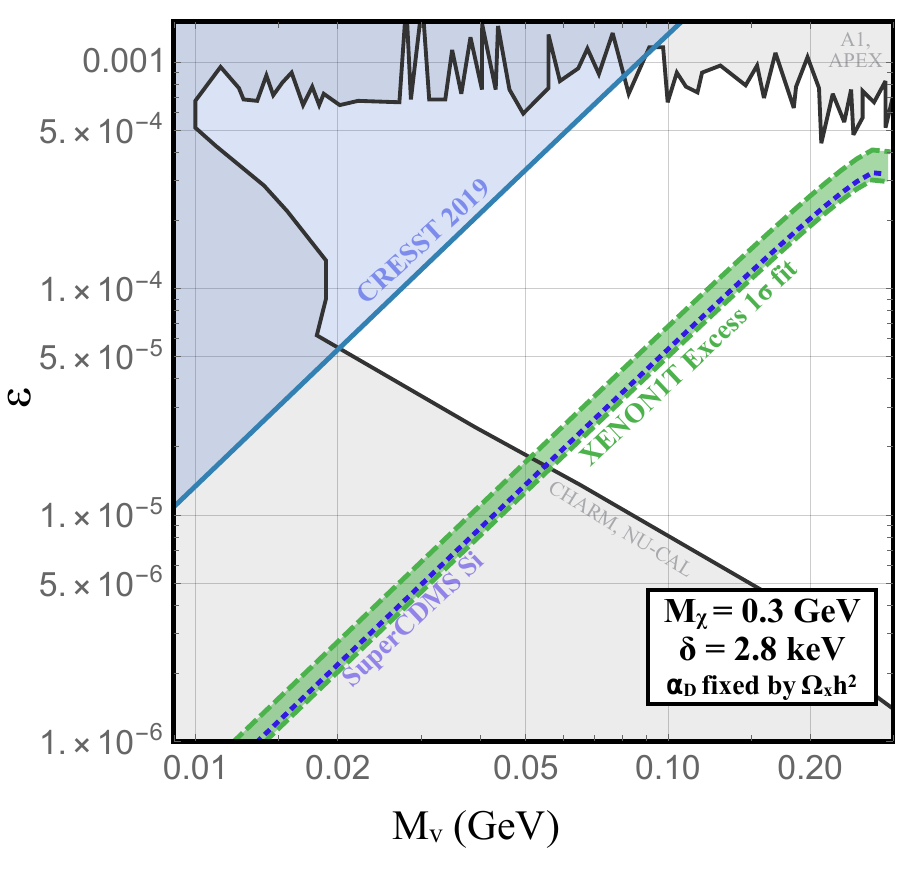}~~~~~~
   \includegraphics[width=0.85\columnwidth]{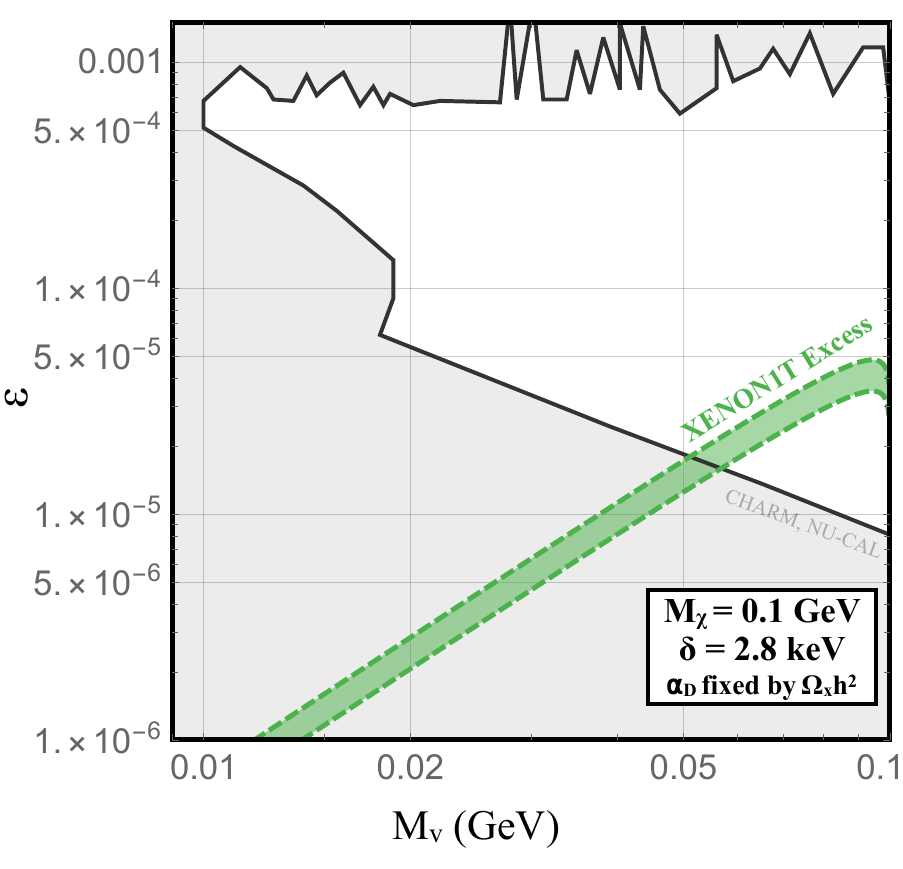}
\caption{This figure shows Dirac fermion DM parameters that provide for the observed DM abundance through thermal freeze-out processes in the early universe, while simultaneously accounting for the observed excess of electron recoil events at XENON1T. Throughout, $\alpha_D$ has been fixed to yield the observed cosmological abundance of DM, according to Eq.~\eqref{eq:adrelic}. The green region enclosed by dashed lines shows the $1 \sigma$ best fit inelastic downscattering rate matching the observed XENON1T excess. The electron scattering cross-sections corresponding to these parameters are shown in Figure \ref{fig:eventrate}. The mass splitting between Dirac fermion component states $\delta$ and DM mass $M_\chi$ are indicated. Constraints on dark photons are shown \cite{Abrahamyan:2011gv,Essig:2013lka,Goudzovski:2014rwa,Merkel:2014avp,Batell:2014mga,Marsicano:2018glj} (gray) alongside CRESST DM-nucleon scattering bounds \cite{Abdelhameed:2019hmk} (blue), and SuperCDMS Ge/Si projections \cite{Agnese:2016cpb} (dotted).}
   \label{fig:param}
\end{figure*}    

In the presence of a heavier DM state $\chi_2$ and a lighter state $\chi_1$, three possible DM-electron scattering processes may take place in a DM detector: (a) elastic: $\chi_{1(2)}+e\rightarrow \chi_{1(2)}+e$, (b) endothermic: $\chi_{1}+e\rightarrow \chi_{2}+e$, (c) exothermic: $\chi_{2}+e\rightarrow \chi_{1}+e$. The typical recoil energy in process (a) and (b) is $\mu_{\chi e}v^2\sim$~eV, which is much smaller than the recoil energy required to explain the XENON1T excess. For the DM model we consider here, (c) exothermic electron scattering would be the only process detected at XENON1T. In this scenario, the electron recoil energy is mainly extracted from the down scattering of $\chi_2$. From energy conservation we can solve for the momentum transfer $q=|\vec{q}|$
\begin{equation}
    q=k\cos\beta\pm\sqrt{k^2\cos^2\beta+2M_\chi(\delta -E_R)}\,,
\end{equation}
where $\cos\beta$ accounts for the scattering angle between the momentum $\vec{k}$ of $\chi_2$ and $\vec{q}$. The maximum and minimum momentum transfer are
\begin{equation}
    q_{\min,\max}=|k\mp\sqrt{k^2+2M_\chi(\delta -E_R)}|\,,
\end{equation}
and the minimum velocity 
\begin{equation}
    v_{\min}^2=\max\left\{\dfrac{2}{M_\chi}(E_R-\delta ),\,0\right\}.
    \label{eq:vmin}
\end{equation}

Following~\cite{Roberts:2016xfw,Roberts:2019chv} the velocity-averaged differential cross section in exothermic DM scattering reads
\begin{equation}
    \dfrac{d\langle \sigma v\rangle}{dE_R}=\int_{v_{\min}}^{v_{\max}}\dfrac{dv}{v}f(v)\int_{q_{\min}}^{q_{\max}}\dfrac{\sigma_e}{2m_e}a_0^2qdqK(E_R,q)\,,
    \label{eq:dsigmavdE}
\end{equation}
where $a_0=1/(m_e\alpha)$ denotes the Bohr radius with the fine structure constant $\alpha\simeq 1/137$, and $K(E_R,q)$ is the atomic ionization factor outlined in~\cite{Roberts:2016xfw,Roberts:2019chv}. For $E_R\sim 2$~keV, the characteristic momentum transfer $q$ is about tens of keV, which corresponds to $K\sim 0.1$. We take a standard Boltzmann DM velocity distribution $f(v)$, where the angular part has been integrated over. We assume the Earth velocity $v_e=240$~km/s and the escape velocity $v_{esc}=600$~km/s. The maximum velocity of DM is then $v_{\max}=v_e+v_{esc}$. In the limit where dark photon mass $M_V$ is much larger than the momentum transfer, the scattering cross section takes the form~\cite{Pospelov:2007mp,Batell:2009vb}
\begin{equation}
    \sigma_e=\dfrac{16\pi\epsilon^2\alpha\alpha_D\mu_{\chi e}^2}{M_V^4}.\,,
    \label{eq:xsheavy}
\end{equation}

The electron recoil energy will be smeared by the detector resolution, which to a good approximation can be modeled by~\cite{Aprile:2020yad}
\begin{equation}
    \dfrac{\sigma_\mathrm{det}}{E_R}=\dfrac{a}{\sqrt{E_R/\mathrm{keV}}}+b\,,
    \label{eq:sigmadet}
\end{equation}
where $a=0.3171\pm0.0065$ and $b=0.0015\pm0.0002$. This gives a resolution of 23\% at 2~keV. We take the Gaussian resolution function 
\begin{equation}
    Res(E,E_R)=\dfrac{1}{\sqrt{2\pi\sigma_\mathrm{det}^2}}e^{-\frac{(E-E_R)^2}{\sigma_\mathrm{det}^2}}\alpha(E)\,,
\end{equation}
which incorporates the efficiency $\alpha(E_R)$ reported in~\cite{Aprile:2020tmw} and can be convoluted with the velocity-averaged cross section in Eq.~\eqref{eq:dsigmavdE} to produce the DM detection rate in the XENON1T detector
\begin{equation}
    \dfrac{dR}{dE}=N_T\dfrac{\rho_{\chi_2}}{M_\chi}\int\dfrac{d\langle \sigma v\rangle}{dE_R}Res(E,E_R)dE_R\,,
\end{equation}
where $N_T\simeq 4.2\times 10^{27}/$ton is the number of Xenon atoms in the detector, and $\rho_{\chi_2}$ is the energy density of $\chi_2$. As detailed in the preceding section, it is safe to assume that half the DM particles are in the excited $\chi_2$ state for the model parameters we are interested in, in which case $\rho_{\chi_2} \simeq 0.15 ~{\rm GeV/cm^3}$.

We show the expected event rate from exothermic scattering in Figure~\ref{fig:eventrate} for a best-fit inelastic mass splitting $\delta=2.8$ keV. Regardless of $M_\chi$, the scattering rate exhibits a sharp peak around $\delta $ before detector resolution smearing. The rate drops abruptly as $E_R>\delta $ for $M_\chi=0.1$~GeV, due to a relatively large $v_{\min}$ as can be understood from Eq.~\eqref{eq:vmin}. Therefore the recoil energy peak for 1~GeV DM tends to be more symmetric. However, this difference in the recoil spectra should not be noticeable in practice, since the recoil energy spectra are appreciably smeared by the detector resolution as given by Eq.~\eqref{eq:sigmadet}. We see from the upper panel of Figure~\ref{fig:eventrate} the smeared scattering spectrum with background can describe the XENON1T data quite well. We have fit the XENON1T data~\cite{Aprile:2020tmw} in the 1~keV-30~keV range by fixing $\rho_{\chi_2}=\rho_\mathrm{DM}/2$ and varying $\sigma_e$. We assume $3\%$ Gaussian error on the efficiency $\alpha(E)$ consistent with~\cite{Boehm:2020ltd}. Although small $M_\chi\sim 10$~MeV prefers slightly larger $\delta$, we fix $\delta =2.8$~keV in the analysis.

The $1 \sigma$ best fit exothermic electron scattering cross section is shown in Figure~\ref{fig:eventrate}. Because of the detector resolution and kinematic uniformity of exothermic scattering detailed above, the fit does not change appreciably with DM mass: \mbox{$\Delta \chi^2=\chi^2_\mathrm{WIMP+bkgd}-\chi^2_\mathrm{bkgd}=-9.8 \rightarrow -10.6$} when $M_\chi$ increases from 10~MeV $\rightarrow$ 1GeV. 

In Figure \ref{fig:param} we show parameter space where DM is produced in the correct relic abundance in the early universe, and which also predicts an excess of events at XENON1T through exothermic DM-electron scattering. Besides scattering with electrons, dark photon mediated DM may also scatter with nuclei, predominantly through scattering with protons. The per-nucleon scattering cross-section against a nucleus with nucleon number $A$ and proton number $Z$ is
\begin{align}
    \sigma_n=\dfrac{16\pi\epsilon^2\alpha\alpha_D\mu_{\chi n}^2}{M_V^4} \left( \frac{Z}{A}\right)^2,
\end{align}
where $\mu_{\chi n}$ is the DM-nucleon reduced mass. For most low mass nuclei, $Z/A = 0.5$, including oxygen at CRESST \cite{Abdelhameed:2019hmk}, which sets a leading bound on sub-GeV mass DM-nucleon elastic scattering, which we have shown in Figure \ref{fig:param}. At present, CRESST's elastic scattering bound provides the most stringent constraint on nuclear scattering for this model, since the exothermic $\mu_{\chi n} \delta/M_N \sim 0.1$ keV recoil energy contribution is comparable to elastic recoil energies for sub-GeV DM at CRESST. An exothermic reanalysis of CRESST recoil data might provide a slightly tighter bound. This would require properly modeling CRESST's low energy recoil backgrounds. For $M_\chi = 0.1$ GeV, a weaker bound on DM-nucleon scattering can be derived using the Migdal effect and results from the XENON1T experiment \cite{Essig:2019xkx}. However, this constraint on $\sigma_n$ is too weak to appear in Figure \ref{fig:param}.

\section{Discussion}

We have studied a specific model of inelastic dark photon mediated dark matter, and found that a sub-GeV Dirac fermion coupled to a lighter sub-GeV mass dark photon could account for the XENON1T excess, while simultaneously predicting the correct relic abundance of dark matter through freeze-out annihilation in the early universe. A crucial feature of this model is a few keV mass splitting between the component Dirac states, resulting in exothermic electron scattering events at XENON1T.

There are many avenues for future research. While at present, the dark matter-nucleon cross-section predicted by this model is too weak to be found out at experiments like CRESST, SuperCDMS, and NEWS-G \cite{Abdelhameed:2019hmk,Agnese:2016cpb,Arnaud:2017bjh}, these experiments are projected to reach sensitivities that should test this model for Dirac fermion masses down to 0.3 GeV, as shown in Figure \ref{fig:param}. In addition, as more electron recoil events are collected and detector resolution improves at xenon experiments like XENON, PandaX, and LZ \cite{Aprile:2015uzo,Cui:2017nnn,Mount:2017qzi}, it should become clear whether the electron recoil spectrum exhibits the sharp peak at a few keV as predicted for exothermic dark photon dark matter in Fig. \ref{fig:eventrate}. We look forward to pursuing these strategies on the path to unveiling the identity of dark matter.

\section*{Acknowledgements}
We thank Fei Gao, Aaron Vincent, and Luc Darme for useful discussions and correspondence. The work of JB, NS is supported by the Natural Sciences and Engineering Research Council of Canada (NSERC). Research at Perimeter Institute is supported in part by the Government of Canada through the Department of Innovation, Science and Economic Development Canada and by the Province of Ontario through the Ministry of Colleges and Universities.

\bibliography{inelastic}
 
\end{document}